\documentclass[%
reprint,
 amsmath,amssymb,
 aps,
]{revtex4-1}
\usepackage{wrapfig}
\usepackage{caption}
\usepackage{subfig}
\usepackage{amssymb}
\usepackage{graphicx}
\usepackage[]{caption}	
\usepackage{amsmath}
\usepackage{multirow}
\usepackage{booktabs}
\usepackage{url}
\usepackage{epstopdf}
\usepackage{tabularx}
\usepackage[T1]{fontenc}
\usepackage[utf8]{inputenc}
\usepackage{floatrow}
\usepackage{dcolumn}
\usepackage{bm}

\input{Qcircuit}

\begin{document}

\title{A Space-Efficient Design for a Reversible Floating Point Adder\\ in Quantum Computing} 



\author{ Trung Duc Nguyen
}
\author{ Rodney Van Meter
}
\affiliation{Keio University, Faculty of Environment and Information Studies \\ 5322 Endo, Fujisawa, Kanagawa, Japan}

\begin{abstract}
Reversible logic has applications in low-power computing and quantum computing. However, there are few existing designs for reversible floating-point adders and none suitable for quantum computation. In this paper we propose a space-efficient reversible floating-point adder, suitable for binary quantum computation, improving the design of Nachtigal et al.~\cite{nachtigal}. Our work focuses on improving the reversible designs of the alignment unit and the normalization unit, which are the most expensive parts. By changing a few elements of the existing algorithm, including the circuit designs of the RLZC (reversible leading zero counter) and converter, we have reduced the cost about 68\%. We also propose fault-tolerant designs for the circuits. The KQ for our fault-tolerant design is almost sixty times as expensive as for a 32-bit fixed-point addition. We note that the floating-point representation makes in-place, truly reversible arithmetic impossible, requiring us to retain both inputs, which limits the sustainability of its use for quantum computation.  

\keywords{reversible circuit, quantum computing, low-power computing, nano technology, IEEE-754 specification} 
\end{abstract}


\maketitle 
\bibliographystyle{plain}

\section{Introduction}

In irreversible systems, erasure of a single bit generates $kT\ln2$ \emph{joules} of heat energy where $k$ is Boltzmann's constant of $1.38\times10^{-23}m^{2}s^{-2}kgK^{-1}$ and $T$ is the absolute temperature of the environment. Based on this observation, Landauer showed that for a reversible computer the energy dissipation is exactly $kT\ln1$ which is equal to zero~\cite{heat}. This means reversible logic has applications in low power computing. Additionally, quantum computing inherently uses reversible computing because most operations in quantum computing are unitary and work as reversible functions. For these reasons, reversible computing is receiving a lot of attention from quantum researchers. 
\par Existing quantum algorithms generally operate on fixed-point numbers, but floating point arithmetic would likely benefit some algorithms~\cite{harrow:lineqs,clader2013quantum,jordan12:qa-qft}. While many fixed point adder designs have been introduced, we are aware of only one design for a floating-point adder, by Nachtigal, Thapliyal and Ranganathan (NTR), and this design is expensive. Our proposed design solves this problem by improving the expensive parts in the NTR design~\cite{nachtigal}. About 68\% of the cost has been eliminated. Moreover, the NTR design as presented leaves many temporary variables in a dirty state, making it unsuitable as-is for quantum computing; our design reduces this number and shows how to compose this design in a fully-reversible setup.
\par A truly reversible circuit generally calculates $ \langle A,B \rangle \xrightarrow{U} \langle A,f(A,B) \rangle $
 where each element of the tuple is a fixed-size register and $U$ is a unitary operation or set of operations that realizes $f(A,B)$. The NTR's circuit actually calculates $ \langle A,B,0,0 \rangle \xrightarrow{U} \langle A,B,A+B,G \rangle$ where $A$, $B$ and $A+B$ are single precision floating point numbers and $G$ is a large amount of ancillary data left in a garbage state. We adapt Bennett's original reversible formulation,

\begin{align}
   \langle A,B,0,0,0 \rangle \xrightarrow{U} & \langle A',B',f(A,B),G,0 \rangle \\\nonumber
   \xrightarrow{CNOT} & \langle A',B',f(A,B),G,f(A,B) \rangle  \\\nonumber
   \xrightarrow{U^\dag} & \langle A,B,0,0,f(A,B) \rangle . 
\label{eqn:Benett}
\end{align}

\par This reduces the garbage output, but cannot solve the fundamental problem that floating point addition is not 1:1, requiring us to retain both inputs as well as the output. Thus, quantum circuits that require many floating point operations may result in unsustainable growth of memory resources.  

\par This paper is divided into six parts. Section 2 reviews reversible logic, evaluation metrics and the IEEE-754 single-precision floating point specification. Section 3 briefly describes the floating-point adder algorithm while Section 4 shows our proposed designs. The comparison between the NTR design and our proposed design and fault-tolerant designs are in Section 5 and 6 respectively. Section 7 concludes. 

\section{Background}

\subsection{Metrics for Evaluating Quantum Circuits}

\begin{figure}[h]
\centering
\subfloat[Unitary matrix]{
$
\begin{pmatrix}
1 & 0 & 0 & 0 & 0 & 0 & 0 & 0\\
0 & 1 & 0 & 0 & 0 & 0 & 0 & 0\\
0 & 0 & 1 & 0 & 0 & 0 & 0 & 0\\
0 & 0 & 0 & 1 & 0 & 0 & 0 & 0\\
0 & 0 & 0 & 0 & 0 & 0 & 0 & 1\\
0 & 0 & 0 & 0 & 0 & 0 & 1 & 0\\
0 & 0 & 0 & 0 & 1 & 0 & 0 & 0\\
0 & 0 & 0 & 0 & 0 & 1 & 0 & 0

\end{pmatrix} 
$

}\hspace{2em}
\subfloat[Barenco decomposition]{
\mbox{
\Qcircuit @C=1em @R=.7em @!R{
	&\lstick{A}  & \qw   &\qw       	&\ctrl{2}   	&\ctrl{1}	&\qw   	&\qw		&\rstick{P=A} \\
	&\lstick{B}  &\qw    &\ctrl{1}  	&\qw 		&\targ    	&\ctrl{1}   	&\qw    &\rstick{Q=A\oplus B}\\
	&\lstick{C}  &\qw   	&\gate{V^\dag} &\gate{V^\dag}  &\qw  	&\gate{V}   &\qw &\rstick{R=AB\oplus C}\\
	}
}
}
\caption{Peres gate}
\label{fig:Peres gate}
\end{figure}
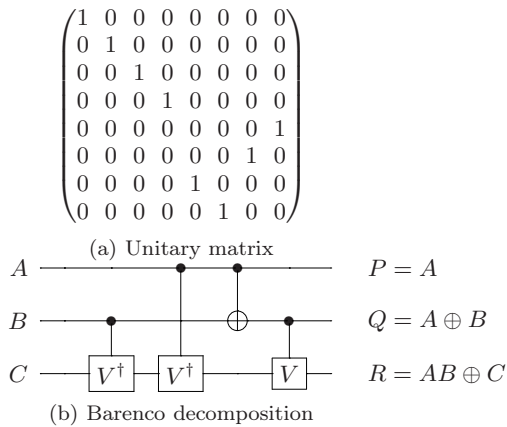

There are various methods to evaluate a circuit but in this paper we adapt Nachtigal's approach of using the quantum cost, the number of constant inputs and the number of garbage outputs. We define the quantum cost as the number of basic gates, while garbage output is the number of unnecessary output qubits which must be cleaned up later. These will be noted as $G$ in the output of our circuits. Constant inputs are qubits with a fixed value, often used to emulate Boolean logic in reversible logic, or as constant values in algorithms. Reversible circuits must always have the same number of inputs and outputs, thus the number of constant plus variable inputs must be equal to the number of useful variable outputs plus garbage. Fig.~\ref{fig:TR gate}(a) shows the unitary operator of the TR gate~\cite{efficient-subtractor} which we use rather than the Peres gate~\cite{peres} that Nachtigal favors, while Fig.~\ref{fig:TR gate}(b) shows the quantum Barenco decomposition~\cite{barenco}. This gate has quantum cost of 4, no constant inputs, and 2 garbage outputs if we only use the third output. The Peres gate's unitary matrix and Barenco decomposition are shown in Fig.~\ref{fig:Peres gate}. The $V$ and $V^{\dag}$ operators, which are unique to quantum computation, behave as follows:
\begin{center}
$
VV=V^{\dag}V^{\dag}=X, 
$
\\
$
VV^{\dag}=V^{\dag}V=I,
$
\end{center}

where $X$ is the Pauli gate corresponding to classical \textsc{not}. The $V$ operator is 

\begin{center}
$
V=\frac{\displaystyle 1}{\displaystyle 2}
\begin{pmatrix}

1+i & 1-i \\
1-i & 1+i

\end{pmatrix}.
$
\end{center}

\begin{figure}[h]

\centering

\subfloat[Unitary matrix]{
$
\begin{pmatrix}
1 & 0 & 0 & 0 & 0 & 0 & 0 & 0\\
0 & 1 & 0 & 0 & 0 & 0 & 0 & 0\\
0 & 0 & 1 & 0 & 0 & 0 & 0 & 0\\
0 & 0 & 0 & 1 & 0 & 0 & 0 & 0\\
0 & 0 & 0 & 0 & 0 & 0 & 1 & 0\\
0 & 0 & 0 & 0 & 0 & 0 & 0 & 1\\
0 & 0 & 0 & 0 & 0 & 1 & 0 & 0\\
0 & 0 & 0 & 0 & 1 & 0 & 0 & 0

\end{pmatrix} 
$

}\hspace{2em}
\subfloat[Barenco decomposition]{
\mbox{
\Qcircuit @C=1em @R=.7em @!R{
	&\lstick{A}  & \qw   &\qw       	&\ctrl{1}   	&\ctrl{2} 	&\qw   	&\qw		&\rstick{P=A} \\
	&\lstick{B}  &\qw    &\ctrl{1}  	&\targ 		&\qw   	&\ctrl{1}   	&\qw    &\rstick{Q=A\oplus B}\\
	&\lstick{C}  &\qw   	&\gate{V^\dag} &\qw  	&\gate{V}   	&\gate{V}   &\qw &\rstick{R=A\overline{B}\oplus C}\\
	}
}

}

\caption{TR gate}
\label{fig:TR gate}
\end{figure}

Quantum computers are far more susceptible to making errors than conventional digital computers, and some method of controlling and correcting those errors will be needed to prevent a quantum computer from making undetected errors in the computation. A device that works effectively even when its elementary components are imperfect is said to be fault-tolerant. 

In fault-tolerant quantum computation~\cite{quantum-fault-tolerant}, we often use another set of gates to build up circuits. The most commonly used set of gates sufficient for universal fault-tolerant quantum computation is the Clifford+$T$ set~\cite{clifford}. Because of this we also use another metric to evaluate a circuit design. The number of $T$ or $T^\dag$ gate is used to calculate quantum cost and $T$-depth, which is the number of steps using a $T$ or $T^\dag$ gate. The reason is because $T$ or $T^\dag$ gates are the most expensive elements in Clifford+$T$ set. The $T$ gate is actually a phase shift gate which modifies the phase of the quantum state without changing the probability of measuring a $\ket{0}$ or $\ket{1}$. The $T$ gate is

\begin{center}
$
T=
\begin{pmatrix}

1 & 0 \\
0 & e^{\frac{i\pi}{4}}

\end{pmatrix}. 
$
\end{center}

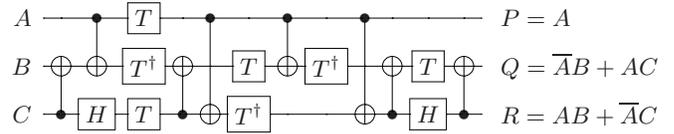
\begin{figure}[h]
\raggedright
\mbox{
\Qcircuit @C=0.3em @R=.5em @!R{
	&\lstick{A}  &\qw   		&\ctrl{1} 	&\gate{T}		&\qw			&\ctrl{2} 	&\qw 		&\ctrl{1}	&\qw 			&\ctrl{2} 	&\qw			&\qw 		&\qw 		&\qw		&\rstick{P=A} \\
	&\lstick{B}  &\targ 	 	&\targ 	 	&\gate{T^\dag}	&\targ 		&\qw 		&\gate{T} 	&\targ	&\gate{T^\dag} 	&\qw			&\targ		&\gate{T} 	&\targ	 	&\qw		&\rstick{Q=\overline{A}B+AC}\\
	&\lstick{C} 	 &\ctrl{-1} 	&\gate{H} 	&\gate{T}		&\ctrl{-1} 	&\targ  		&\gate{T^\dag}  &\qw  	&\qw    			&\targ 		&\ctrl{-1} 	&\gate{H} 	&\ctrl{-1} 	&\qw  	&\rstick{R=AB+\overline{A}C}\\
	}
}

\caption{Decomposition of Fault-Tolerant Fredkin gate.}
\label{fig:ft-Fredkin}
\end{figure}

Fig.~\ref{fig:ft-Fredkin} shows the decomposition of the fault-tolerant Fredkin gate~\cite{fast-synthesis}. In this circuit, the $T$-depth is 4.

\subsection{IEEE-754 Floating-Point}

The IEEE-754 single-precision floating-point format called binary32~\cite{floating-point} uses 32 bits (a 1-bit sign, an 8-bit exponent and a 23-bit mantissa) to represent a floating point number. The sign bit determines the sign of the number, while the exponent is an 8 bit unsigned integer from 0 to 255. The true significand, called the mantissa, includes 23 fraction bits to the right of the binary point and an implicit leading bit (to the left of the binary point) with value 1 unless the exponent is stored with all zeros. Thus only 23 fraction bits of the significand appear in the memory format but the total precision is 24 bits. Suppose that we have a floating-point number with sign bit $s$, exponent $e$ and mantissa $m_{22}m_{21}..m_0$, then the value of the floating point number is calculated as follows:

$(-1)^s\times(1.m_{22}m_{21}...m_0)\times2^{e-127}$ 

\par This format also requires 3 extra bits during computation known as the guard bit, round bit and sticky bit, which must be preserved during the right shifts. The guard and round bits are just two extra bits of precision that are used in calculations. The sticky bit is an indication of what is or could be in less significant bits that are not kept. If a value of 1 ever is shifted into the sticky bit position, that sticky bit remains a 1 ("sticks" at 1), despite further shifts.

\section{Floating-Point Adder Overview}

\begin{figure}[h]
\centering
\includegraphics[width=1\textwidth]{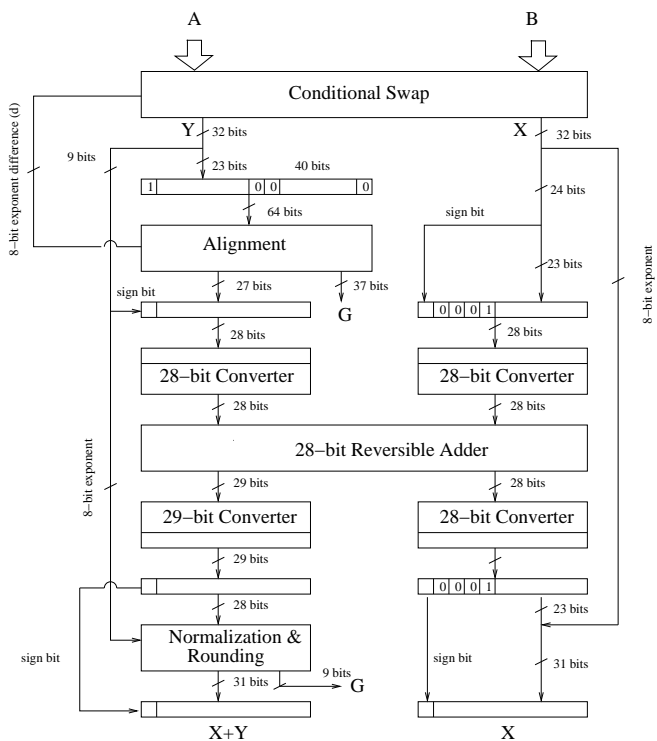}
\caption{Overview of the algorithm for a Floating Point Adder.}
\label{fig:overview}
\end{figure}

In this section the basics of a floating-point adder algorithm will be briefly summarized with attention to the demands of reversibility. Two 32-bit IEEE-754 single-precision floating-point numbers $A$ and $B$ are to be added. Before two numbers can be added, they must be aligned. If the exponents are not equal, the smaller number's exponent is incremented until its exponent reaches the larger number's, in conjunction with shifting the smaller number's mantissa to the right. Once the exponents are equal, the mantissas can be summed. The sum is normalized and rounded at the end. Fig.~\ref{fig:overview} shows the general algorithm adapted to show constant inputs and garbage outputs. The garbage outputs are eventually cleaned by reversing this circuit using Bennett's method.

{\bf Reversible Conditional Swap.} A reversible conditional swap is necessary because we need to figure out which number has the smaller exponent and then input it to the reversible alignment step. If expA $<$ expB (expA stands for the exponent of A) then swap the two numbers, otherwise do nothing. After this step, the number with the smaller exponent always comes out in the Y output, which connects to the reversible barrel shifter~\cite{barrel-shifter} in the next step. 
 
\bigskip

{\bf Reversible Alignment.} We need reversible alignment because we can only add two mantissas when the two exponents are equal, so we need a reversible right shifter to shift the mantissa of the number with smaller exponent, in conjunction with increasing the smaller exponent. The shifted amount is the difference between the two exponents. Because the IEEE-754 floating-point single-precision specification uses an 8-bit exponent, the difference of the exponents is up to 256. 
\par The IEEE-754 specification also requires 3 extra bits as described above. Thus, we need a sticky bit cascade unit to calculate the sticky bit after shifting. The sticky bit is calculated ORing together the $27^{th}$ to $256^{th}$ bits. 

\bigskip

{\bf Two's complement Conversion.} The IEEE-754 specification represents numbers in sign-magnitude format (1 sign bit, 23 mantissa bits). To add two mantissas after the alignment step, the two numbers will be represented in two's complement format. After the addition, the result will be converted back to sign-magnitude format. Thus, we need \emph{sign-magnitude to two's complement} reversible converters and \emph{two's complement to sign-magnitude} reversible converters before and after the addition. The proposed converter will be described later.

\bigskip

{\bf Reversible Addition.} After the \emph{sign-magnitude to two's complement} conversion, the addition is done by a reversible adder which is constructed from 27 RFA (Reversible Full Adder) gates and one RHA (Reversible Half Adder) gate.

\bigskip

{\bf Reversible Normalization and Rounding.} After the addition, the result may have a number of leading zero bits or have one more bit with value of one at the most significant bit (MSB). The normalization is needed to adjust the result so that it conforms to the floating-point number format. In normalization, if a shift is required, it is either a one place right shift or a multiple-place left shift. If the MSB has a value of one, one place of right shift takes place and the 8-bit exponent is passed through a reversible conditional increment unit. Otherwise, one or several places of left shift is needed in conjunction with a corresponding decrement of the 8-bit exponent.

\section{Detailed Design}

In the NTR design, two parts are the main causes of the large quantum cost, the reversible alignment unit and the reversible normalization unit, with 12,312 and 2,009 quantum cost respectively. Our proposed design focuses mainly on these parts, and improves some other parts in smaller ways.   

\subsection{Reversible Conditional Swap}

\par
\begin{figure*}[ht]
\centering
\includegraphics[width=0.8\textwidth]{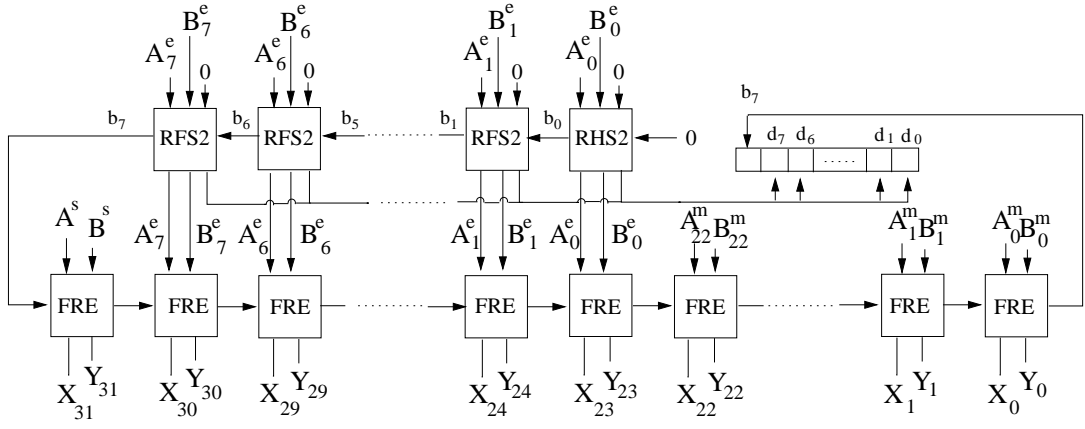}
\caption{Proposed Conditional Swap. The output register d is the difference of the exponents, which later  will be fed into the shift limiter.} 
\label{fig:proposed_swap}
\end{figure*}

\begin{figure}%

\begin{minipage}{2.1in}%
\subfloat[Decomposition]{
\mbox{
\Qcircuit @C=1em @R=.7em @!R{
	&\lstick{A}  & \qw   		&\ctrl{1}   	&\ctrl{2}   	&\qw 	   	&\ctrl{1} &\qw 	&\rstick{P=A} \\
	&\lstick{B}  &\ctrl{1}   	&\targ  		&\qw 		&\ctrl{1}   &\targ    &\qw &\rstick{Q=B}\\
	&\lstick{0}  &\gate{V^\dag}  &\qw   		&\gate{V}   	&\gate{V}   &\qw 	 &\qw 	&\rstick{R=A\overline{B}}\\
	}
}
}
\end{minipage}%
\hspace{7em}
\parbox{1.2in}{
\subfloat[Diagram]{
\includegraphics[width=0.4\textwidth]{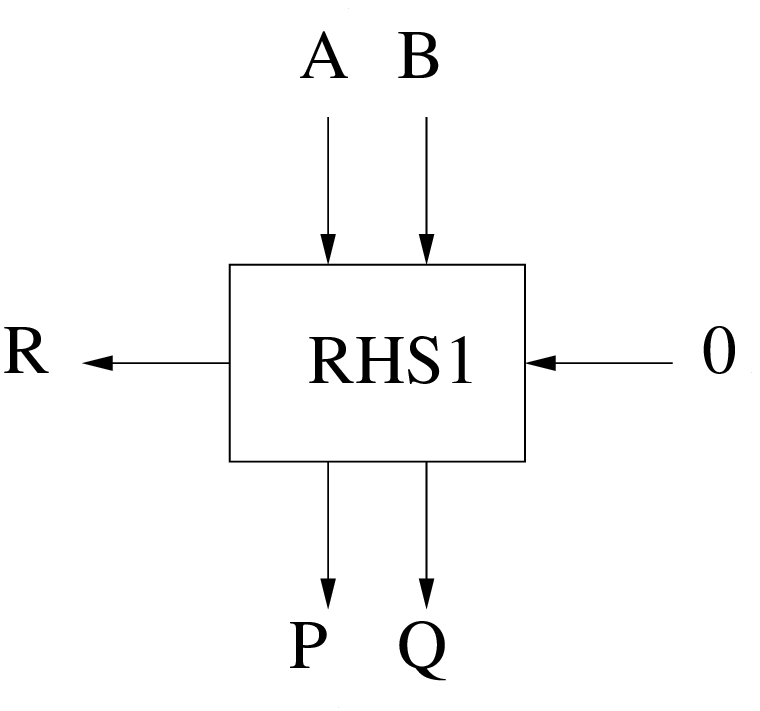}
}
}%

\caption{Diagram and decomposition of the existing Reversible Half Subtractor (RHS1).}%
\label{fig:proposed RHS1}%
\end{figure}

\begin{figure}%

\begin{minipage}{2.1in}%
\subfloat[Decomposition]{
\raggedright
\mbox{
\Qcircuit @C=1em @R=.7em @!R{
	&\lstick{0}  & \qw   		& \qw   		& \qw  		&\qw 	   	& \targ	& \qw	 &\qw 	&\rstick{R=A\oplus{B}} \\
	&\lstick{A}  & \qw   		&\ctrl{1}   	&\ctrl{2}   	&\qw 	   	& \qw 	&\ctrl{1} &\qw 	&\rstick{P=A} \\
	&\lstick{B}  &\ctrl{1}   	&\targ  		&\qw 		&\ctrl{1}   & \ctrl{-2}	&\targ    &\qw &\rstick{Q=B}\\
	&\lstick{0}  &\gate{V^\dag}  &\qw   		&\gate{V}   	&\gate{V}   & \qw	&\qw 	 &\qw 	&\rstick{S=A\overline{B}}\\
	}
}
}
\end{minipage}%
\hspace{7em}
\parbox{1.2in}{
\subfloat[Diagram]{
\includegraphics[width=0.4\textwidth]{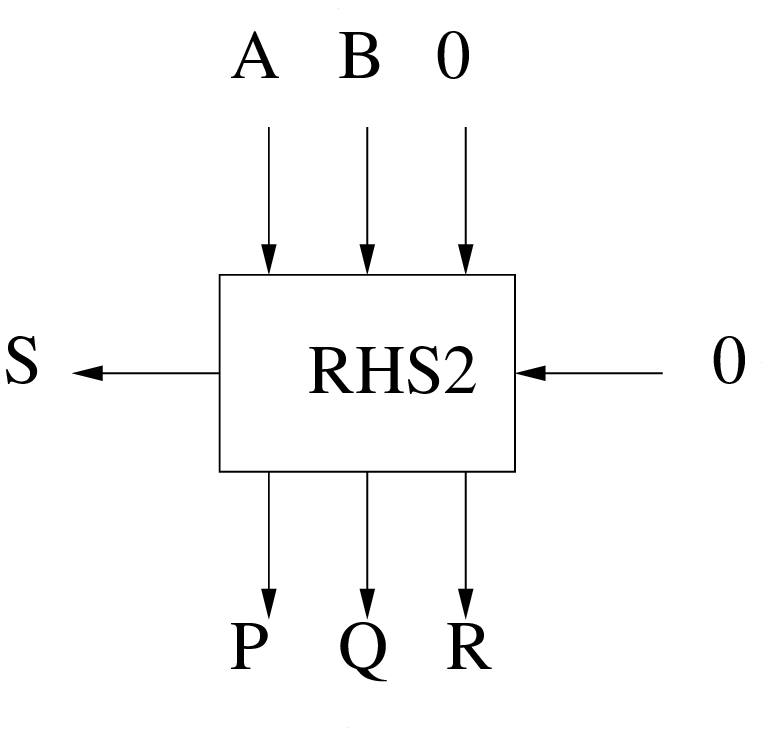}
}
}%

\caption{Diagram and decomposition of our proposed Reversible Half Subtractor (RHS2).}%
\label{fig:proposed RHS2}%
\end{figure}

\begin{figure}%
\begin{minipage}{2.1in}%
\raggedright
\subfloat[Decomposition]{

\mbox{
\Qcircuit @C=0.4em @R=0.3em @!R{
	&\lstick{C}  & \qw         &\qw       &\qw      	&\ctrl{2}   	&\ctrl{3} &\qw  	 &\qw &\rstick{P=C} \\
	&\lstick{B}  & \qw         &\ctrl{1}  &\ctrl{2} 	&\qw      	&\qw   	  &\qw  &\qw &\rstick{Q=B}\\
	&\lstick{A}  &\ctrl{1}     &\targ   	 &\qw 		&\targ      &\qw   	  &\ctrl{1}  &\qw &\rstick{R=A\oplus B\oplus C}\\
	&\lstick{0}  &\gate{V^\dag} &\qw      &\gate{V}  &\qw   		&\gate{V} &\gate{V} 	&\qw &\rstick{S=C(\overline{A\oplus B})\oplus \overline{A}B} \gategroup{1}{5}{4}{6}{.7em}{--}
	} 
}

}
\end{minipage}%
\hspace{7em}
\parbox{1.2in}{
\subfloat[Diagram]{
\includegraphics[scale=0.4]{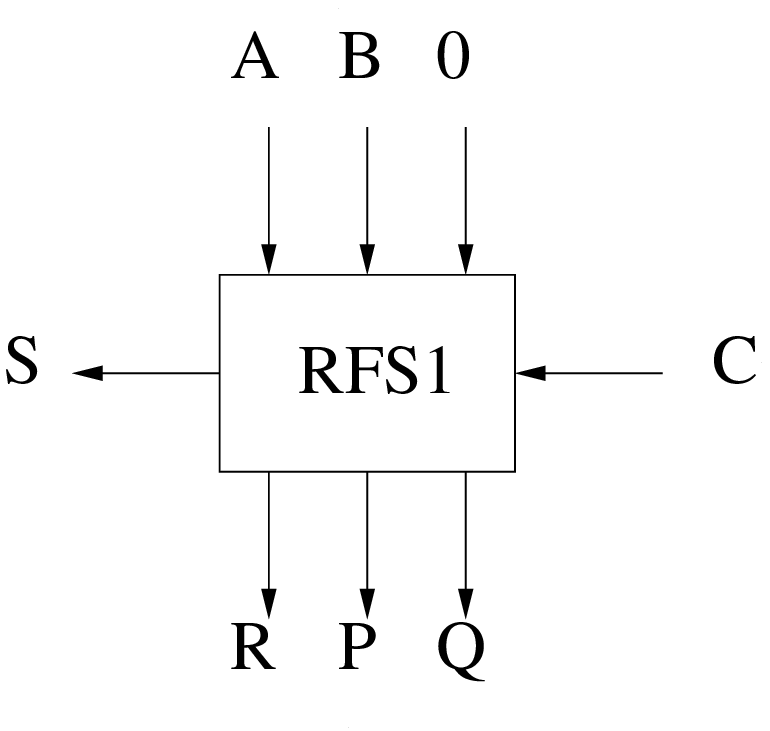}
}
}%

\caption{Diagram and decomposition of the existing Reversible Full Subtractor (RFS1).}%
\label{fig:proposed RFS1}%
\end{figure}

\begin{figure}%

\begin{minipage}{2.1in}%
\raggedright
\subfloat[Decomposition]{
\mbox{
\Qcircuit @C=0.4em @R=0.3em @!R{
	&\lstick{C}  & \qw         &\qw       &\qw      	&\ctrl{2}   	&\ctrl{3} &\targ  &\targ 	 &\qw &\rstick{P=A} \\
	&\lstick{B}  & \qw         &\ctrl{1}  &\ctrl{2} 	&\qw      	&\qw   	  &\ctrl{-1} &\qw    &\qw &\rstick{Q=B}\\
	&\lstick{A}  &\ctrl{1}     &\targ   	 &\qw 		&\targ      &\qw   	  &\ctrl{1} &\ctrl{-2} &\qw &\rstick{R=A\oplus B\oplus C}\\
	&\lstick{0}  &\gate{V^\dag} &\qw      &\gate{V}  &\qw   		&\gate{V} &\gate{V} &\qw 	&\qw &\rstick{S=C(\overline{A\oplus B})\oplus \overline{A}B} \gategroup{1}{5}{4}{6}{.7em}{--}
	} 
}

}
\end{minipage}%
\hspace{7em}
\parbox{1.2in}{
\subfloat[Diagram]{
\includegraphics[scale=0.4]{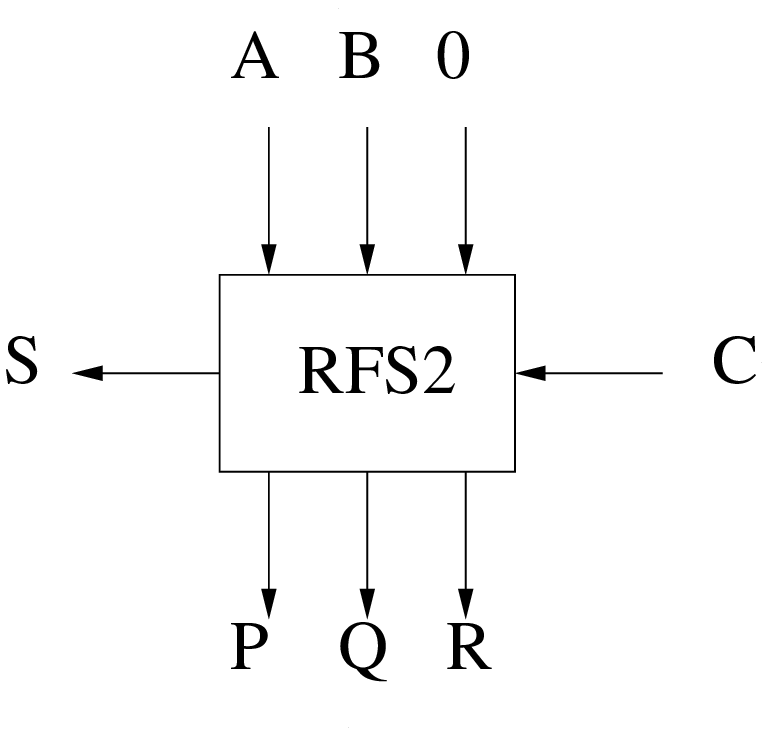}
}
}%

\caption{Diagram and decomposition of our proposed Reversible Full Subtractor (RFS2).}%
\label{fig:proposed RFS2}%
\end{figure}

 In our proposed design, the actual swap is done by a bank of 32 Fredkin gates but we use 7 RFS (reversible full subtractor) and one RHS (reversible half subtractor) to construct the reversible subtractor as shown in Fig.~\ref{fig:proposed_swap}. In the existing design of RHS and RFS~\cite{new-subtractor} in Fig.~\ref{fig:proposed RHS1} and Fig.~\ref{fig:proposed RFS1}, the quantum costs are 4 and 6 respectively. To reduce the garbage output and reuse them we proposed the new design in Fig.~\ref{fig:proposed RHS2} and Fig.~\ref{fig:proposed RFS2}. In Fig.~\ref{fig:proposed_swap}, the $A_i^e$ and $B_i^e$ note the exponent bits of the two floating-point numbers, while $A_i^m$ and $B_i^m$ are the mantissa bits and $A^s$ and $B^s$ are sign bits. The borrow bits $b_i$ from previous RHS or RFS are passed to the next RFS. The high order output bit $b_7$ will be used as a conditional signal for swap and the difference which is output in two's complement format $b_7d_7..d_0$ will be used for the alignment unit as the shifted amount after conversion to sign-magnitude format. 

\subsection{Reversible Alignment} 

\par Although the difference in the exponents may be up to 256, we observe that the mantissa with guard bit and round bit is only 26 bits. Thus, when the difference is greater than 26, the smaller number is effectively discarded and the result is simply the larger number. Thus, we propose using a shift limiter unit, which outputs the smaller of the exponent difference and 26. This output is used to control the barrel shifter. We only need a reversible barrel shifter for 50 input bits and 23 reversible OR gates for the reversible sticky bit cascade unit. We choose to use a (64, 6) reversible barrel shifter~\cite{efficient-barrel}. 

\par One difficulty is the floating point sticky bit, which is the OR of all of the bits shifted past it. Calculating an OR function in reversible logic is difficult, requiring us to keep additional ancillae. In our proposed design, we use TR gates as in Fig.~\ref{fig:TR gate}, first applying a NOT gate to the $A$ input and setting the $C$ input to the constant value 1. The OR operation between $A$ and $B$ input comes out in the $R$ output.

\begin{figure*}[ht]
\centering
\includegraphics[width=0.7\textwidth]{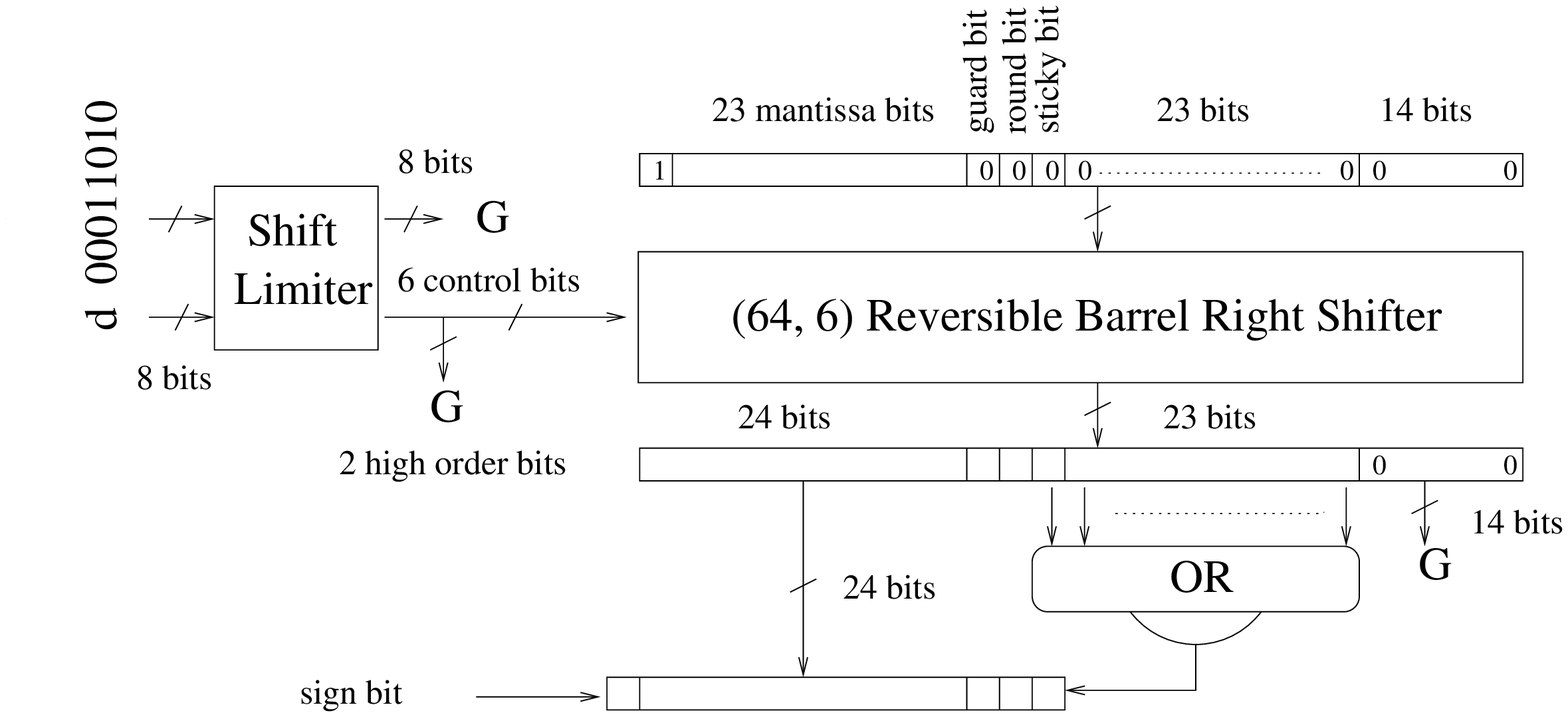}
\caption{Proposed design for Reversible Alignment Unit. The constant value in the top left input is the number 26, the upper limit for our shift distance. $d$ is the exponent's difference.}
\label{fig:proposed_align}
\end{figure*}

Fig.~\ref{fig:proposed_align} shows the proposed design for the alignment unit at the block level.

\par
\begin{figure}[h]
\centering
\includegraphics[width=1\textwidth]{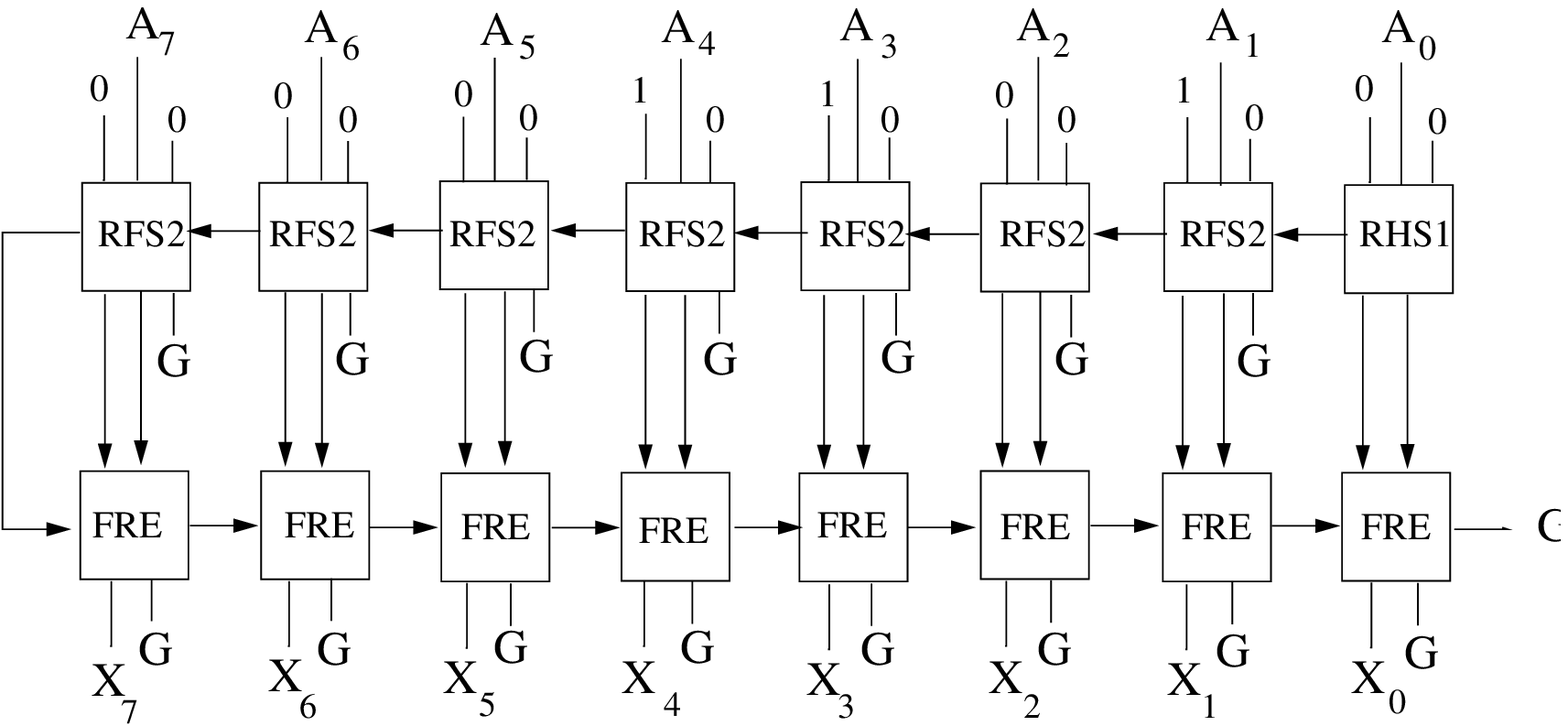}
\caption{The shift limiter works as a reversible conditional swap.}
\label{fig:shift_limiter}
\end{figure}

 The shift limiter works in the same way as conditional swap at the first step except that only 8 bits are swapped instead of 32 bits. It is also slightly different in that we only need the smaller of the number 26 and the exponent's difference so we don't need to use the design of RHS2 in Fig.~\ref{fig:proposed RHS2}, which is used to reduce the number of garbage outputs. Instead, we use the RHS1 in Fig.~\ref{fig:proposed RHS1}. The design of the shift limiter is shown in Fig.~\ref{fig:shift_limiter}.

\subsection{Reversible Converter}

\begin{figure}[h]
\centering
\begin{minipage}{1.2in}
\subfloat[Diagram]{
\includegraphics[scale=0.6]{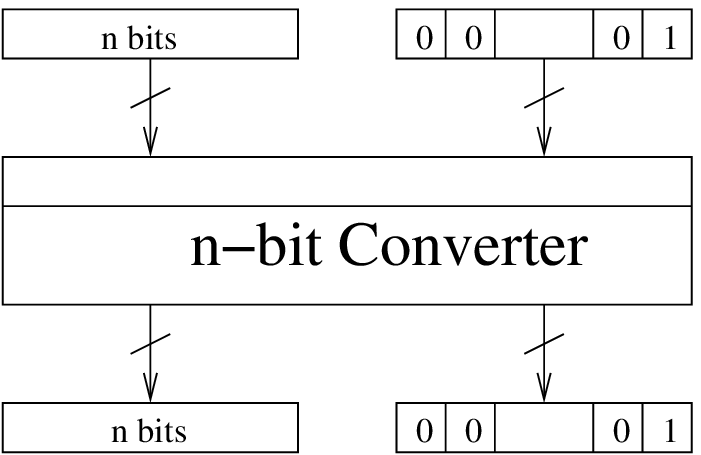}
}
\end{minipage}
\subfloat[Proposed Converter in details]{
\includegraphics[scale=0.55]{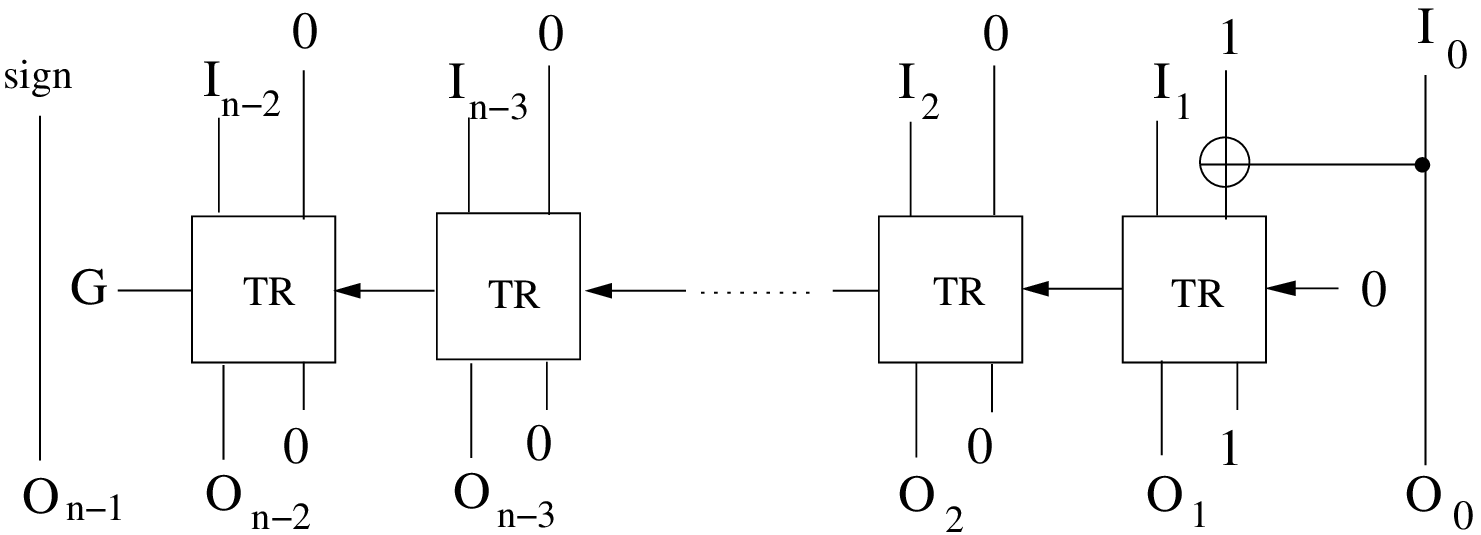}
}
\caption{Proposed design for Reversible $n$-bit Converter.}
\label{fig:proposed_converter}
\end{figure}

Suppose that we have a $n$-bit sign-magnitude number. The algorithm for converting a sign-magnitude number to two's complement is to invert all the bits and add 1. To add 1 to the inverted bits we only need RHA gates instead of RFA gates combined with RHA gate, because we just need to add the carry bit of the previous operation. We realized that one TR gate can do both at the same time. In addition, we observe that after conversion the least significant bit (LSB) does not change and the carry bit for the next bit is the inverse of the LSB. Therefore, we only need 1 CNOT gate and $n-2$ TR gates.  Fig.~\ref{fig:proposed_converter} shows the new design and its diagram.

In the conversion step, we use a total four converters: two \emph{28-bit sign-magnitude to two's complement} converters for pre-addtion, one \emph{28-bit two's complement to sign-magnitude} converter and one \emph{29-bit two's complement to sign-magnitude} for post-addition. We use one \emph{29-bit two's complement to sign-magnitude} to add one more high order bit to avoid overflow during the addition. See Fig.~\ref{fig:overview} for details. 

\subsection{Reversible Normalization and Rounding}

\par One of the problems is how to calculate the left shift amount (if needed). The left shift is the number of leading zero bits. The NTR design used a RLZCU (reversible leading zero counter unit), which requires $n-1$ gates of RLZC (reversible leading zero counter) for $n$-bit input. However, the NTR design is not efficient and we propose a novel reversible leading zero counter design. 
\par The output of reversible normalization is an array of 32 bits, but only the first 23 bits are needed. We used a \emph{round toward zero} rounding algorithm. Thus, we just need to make the first bit and last bits become garbage outputs and no quantum gate is needed. 

\par
\begin{figure}%

\parbox{2in}{
\subfloat[Decomposition]{
\mbox{
\Qcircuit @C=0.2em @R=0.3em @!R{
	&\lstick{A}     &\qw       &\qw      &\qw   	&\qw  &\ctrl{3} &\ctrl{4} 	&\targ 	&\qw 	&\ctrl{4} 			&\targ &\qw &\rstick{A} \\
	&\lstick{B}  & \targ    &\ctrl{1}   	&\ctrl{2} &\qw      &\qw   	&\qw      	&\qw 	  	&\targ  			&\qw &\qw &\qw &\rstick{B}\\
	&\lstick{C}   &\ctrl{1}   &\targ  &\qw  &\ctrl{1}      &\qw   	&\qw      	&\qw 		&\qw  &\qw	&\qw &\qw &\rstick{G}\\
	&\lstick{0}     &\gate{V^\dag}      &\qw      &\gate{V}   	&\gate{V} &\ctrl{3} &\qw &\ctrl{-3} &\ctrl{1} 		&\qw 	&\ctrl{-3}	  		&\qw &\rstick{G}\\
	&\lstick{1}    &\qw        &\qw      &\qw   	&\qw &\qw      &\gate{V^\dag}     &\qw 	&\gate{V}	&\gate{V}  &\qw  &\qw &\rstick{A+B+C}\\
	&\lstick{D}     &\qw        &\qw      &\qw   	&\qw &\qw       &\ctrl{2}	&\targ  		&\qw	 &\ctrl{2}  &\qw  &\qw &\rstick{G}\\
	&\lstick{1}   &\qw        &\qw      &\qw   	&\qw &\targ   &\qw   	&\ctrl{-1}  	&\ctrl{1}     			&\qw &\qw &\qw &\rstick{G}\\
	&\lstick{1}    &\qw        &\qw      &\qw   	&\qw     &\qw   	&\gate{V^\dag} &\qw   &\gate{V} 			&\gate{V} &\qw &\qw  &\rstick{D+AB'C'}\\ 
	} 
}
}
}%
\hspace{2em}
\begin{minipage}{1.2in}%
\subfloat[Diagram]{
\includegraphics[scale=0.4]{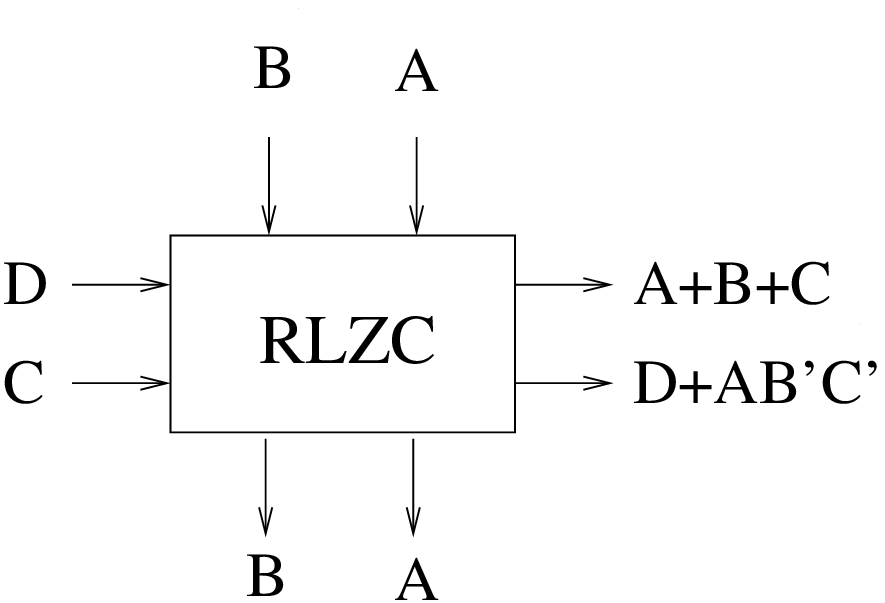}
}

\end{minipage}%
\caption{Diagram and our proposed design for RLZC.}%
\label{fig:proposed_RLZC}%
\end{figure}

 Our proposed design is constructed by using one Toffoli gate, 3 TR gates, 2 NOT gates and 1 CNOT gate as shown in Fig.~\ref{fig:proposed_RLZC}. Because each basic gate is counted as quantum cost 1, this design has quantum cost 20, garbage output 4 and constant input 4, respectively, and even reuses mantissa bits to reduce the number of garbage outputs.

\par
\begin{figure*}[ht]
\centering
\includegraphics[width=0.8\textwidth]{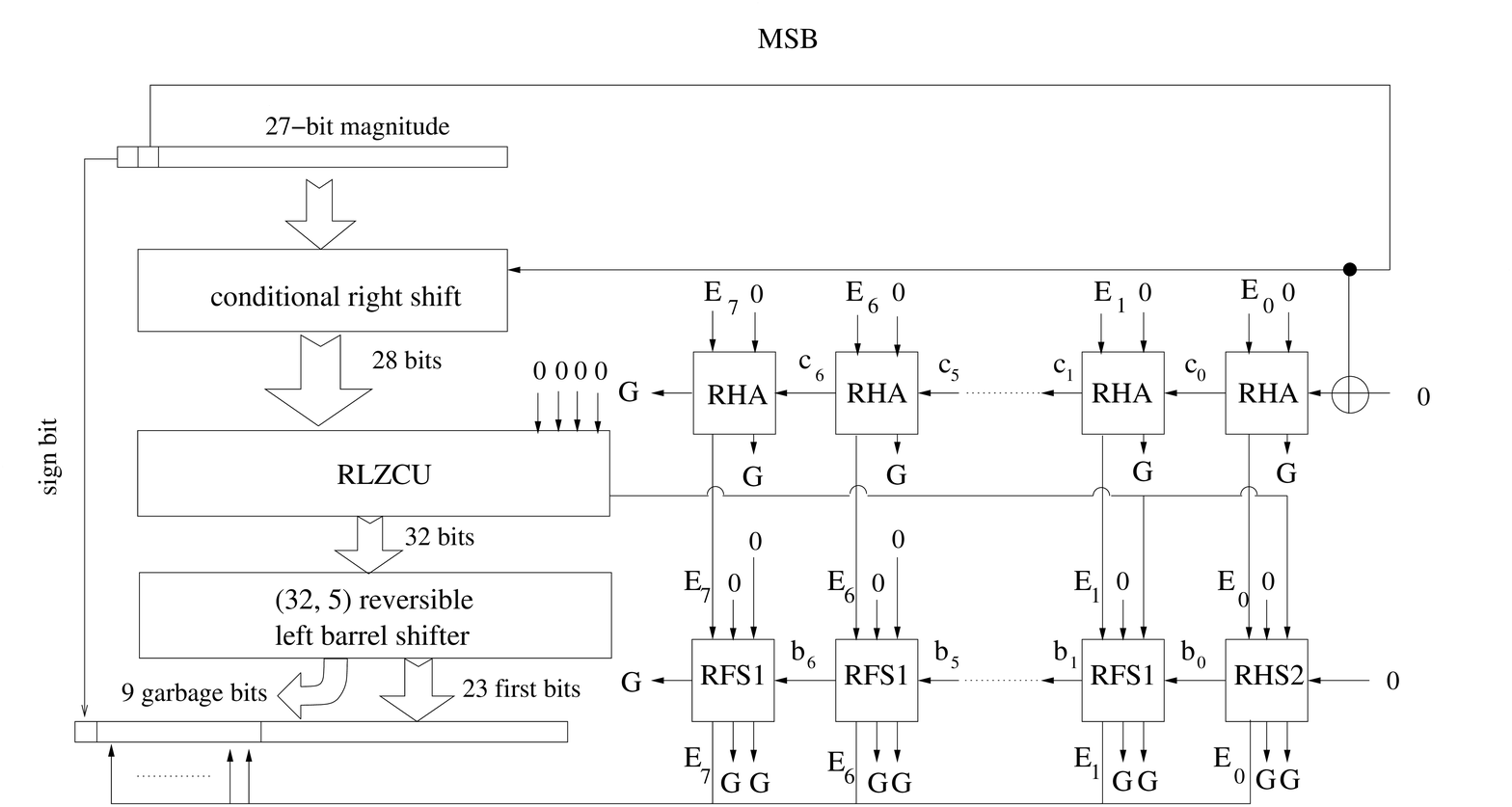}
\caption{Reversible Normalization.}
\label{fig:normalization} 
\end{figure*}

 The shift is in conjunction with exponent addition or subtraction, thus, for these operations we use RHA gates, which are constructed from Peres gates. RFS gates are used to do the subtraction (if needed). We use the design in Fig.~\ref{fig:proposed RFS1} for RFS gates and the design in Fig.~\ref{fig:proposed RHS2} for RHS gate. The carry bits $c_i$ and borrow bits $b_i$ are passed to the next RHA or RFS gates as shown in Fig.~\ref{fig:normalization}. Because the RLZCU has a 5-bit output for 32-bit input while the input for RFS gates needs 8 bits, we add 3 zero bits as the high order bits. Because the conditional right shifter is just a one place shift, we use the (28, 1) reversible barrel shifter described in~\cite{barrel-shifter}. 

\par
\begin{figure}[h]
\centering
\includegraphics[width=0.9\textwidth]{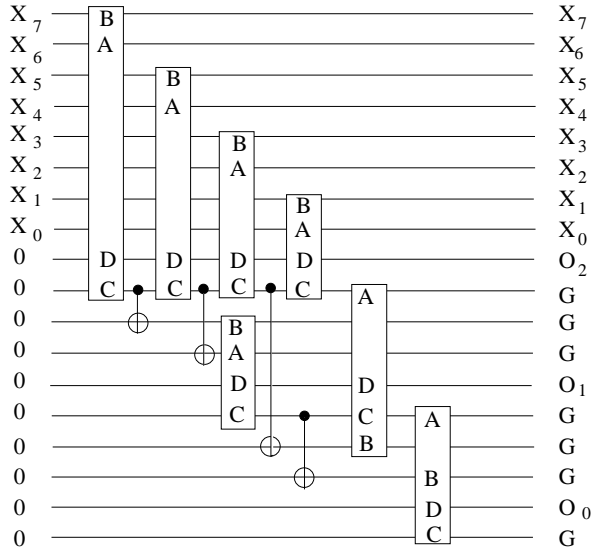}
\caption{An example of using RLZCU for 8 input bits.}
\label{fig:RLZC_example}
\end{figure}

 Fig.~\ref{fig:RLZC_example} shows an example of using RLZCU for an 8-bit input. Suppose that the 8-bit input $X_7X_6...X_0$ is $00111010$. The output $010$ is produced on the bits $O_2O_1O_0$. 

\section{Comparison with the NTR design}

The NTR design is ambiguous in some places, and the evaluation apparently contains several errors. This section describes in details the difference between our proposed design and NTR design.

\subsection{Reversible Conditional Swap}

In the NTR design, the authors proposed using 9 HNG gates~\cite{HNG} as a reversible subtractor. The actual swap is done by a bank of 32 Fredkin gates~\cite{fredkin} controlled by the subtractor's high order output bit. 
While the NTR design includes fanout of exponent bits and produces garbage output, our design tries to reduce the quantum cost, garbage output and fanout by reusing the exponent bits after subtraction. In our proposed design, we use 7 RFS gates and 1 RHS gate to construct the reversible subtractor. The evaluation and comparison is shown in Table~\ref{tab:RLZC}.

\subsection{Reversible Alignment}

A (256, 8) reversible barrel shifter (256 input bits, 8 control lines)~\cite{efficient-barrel} is used in the NTR design. This is one of the key reasons for the large quantum cost, constant input and garbage output of the NTR design. With our proposed design, we only use the (64, 6) reversible barrel shifter which reduces the cost significantly.

\par Nachtigal, Thapliyal and Ranganathan calculated a quantum cost of 4,632 for the alignment unit, using a Fredkin gate cost of 1 in their design. Assigning a cost of 5 to the Fredkin gate, we find a cost of 12,312 for this unit. According to~\cite{efficient-barrel} the number of Fredkin gates and Feynman gates for a (256, 8) reversible barrel shifter should be 1,920 and 1,792 respectively. The authors apparently took the sum of these numbers, ignoring the Fredkin gate's non-unit cost, and arrived at a cost of 3,712, which we believe substantially understates the true cost of their design. 

\par Using Peres gates to make a reversible OR gate, as in the NTR design, will have a quantum cost of at least 6, while our proposed design only requires 5. 

\subsection{Reversible Converter}

The NTR design used $n-1$ Peres gates~\cite{peres} combined with $n-2$ NOT gates for both reversing bits and RHA gates, but we realized that one TR gate can do both at the same time. Our proposed design eliminates the NOT gates. We also save one more TR gate compare to the NTR design. The number of garbage ouput of our proposed design is reduced to 1 because most of the ancillae used in one converter can be reused in another conversions. 

\par Table~\ref{tab:converter} shows a comparison between the NTR reversible converter and our proposed design for $n$-bit input. For the total evaluation of every conversion used in floating-point adder, see Table~\ref{tab:total_compare}. Note that our calculation for the conversions of our proposed design's cost differs from the authors of NTR design, who seem not to have included the cost of conversion for the exponent's difference after conditional swap. While the NTR design only uses 3 converters and leaves one operator of the addition in a dirty state, we use one more \emph{two's complement to sign-magnitude} reversible converter to avoid that problem.

\begin{table}
\centering
\scalebox{0.8}{
\begin{tabular}{ |l|c|c|c| }
\hline
&\textbf{Quantum Cost} &\textbf{Garbage Ouput} &\textbf{Constant Input}\\
\hline
NTR Design & $5n-4$ & $n$  & $n$\\ \hline
Proposed Design & $4n-7$  & $1$  & $n-1 $\\ \hline
\end{tabular}
}
\\
\caption{NTR Converter vs Proposed Converter.}
\label{tab:converter}
\end{table}

\subsection{Reversible Normalization and Rounding}
By reducing the cost for each RLZC, our proposed design has substantially reduced the cost for the whole Reversible Leading Zero Counter Unit which is constructed from 31 RLZCs. Table~\ref{tab:RLZC} compares the NTR design and our proposed design for RLZC.

\begin{table}
\centering
\scalebox{0.8}{
\begin{tabular}{ |l|c|c|c| }
\hline
&\textbf{Quantum Cost} &\textbf{Garbage Ouput} &\textbf{Constant Input}\\
\hline
NTR Design & 27 & 10  & 8\\ \hline
Proposed Design & 20  & 4  & 4 \\ \hline
Reduction Ratio & 26\% & 60\% & 50\% \\\hline 
\end{tabular}
}
\\
\caption{NTR RLZC vs Proposed RLZC.}
\label{tab:RLZC}
\end{table}

\subsection{Overall Comparison}
We have reduced the Quantum Cost by 68\%, the Garbage Output by 72\% and the Constant Input by 71.5\%. Table~\ref{tab:total_compare} compares the NTR design and our proposed design in quantum cost, garbage output and constant input. 

\begin{table}
\centering
\begin{tabular}{|c|c|c|c|c|c|c|}
\hline
\textbf{Stage} & \multicolumn{3}{|c|}{\textbf{NTR Design}} & \multicolumn{3}{|c|}{\textbf{Proposed Design}} \\
\cline{2-7}
 & QC & GO & CI & QC & GO & CI  \\ \hline 
 Swap & 238 & 19 & 27 & 220 & 0 & 9 \\ \hline
 Alignment & 12312 & 2260 & 2022 & 2295 & 388 & 359 \\ \hline
 Addition & 166 & 55 & 28 & 166 & 55 & 28  \\ \hline
 Conversion & 454 & 94 & 94 & 450 & 56 & 55 \\ \hline
 Normalization & 2009 & 498 & 484 & 1742 & 313 &306 \\ \hline
 Rounding & 0 & 9 & 0 & 0 & 9 & 0 \\ \hline 
 Total & 15179 & 2935 & 2655 & 4873 & 824 & 757 \\ \hline 
\end{tabular}
\\
\caption{NTR Design vs Proposed Design.}
\label{tab:total_compare}
\end{table} 

\section{Fault-Tolerant Design}

In this paper, we have seen a lot of designs using controlled-$V$ and controlled-$V^\dag$ gates. But in quantum computing, the direct fault-tolerant implementation of controlled-$V$ and controlled-$V^\dag$ gates is very hard. To make our design implementable in quantum computing we need to use fault-tolerant forms of the TR gate, Peres gate, Fredkin gate and Toffoli gate. In this section we introduce these fault-tolerant circuit architectures. 

\par Recently, M. Amy et al.~\cite{fast-synthesis} have introduced several depth-optimal decompositions of the Toffoli gate, Peres gate and TR gate. Fig.~\ref{fig:fault-tolerant Toffoli, Peres, TR} shows the fault-tolerant designs of these gates.

\begin{figure}[h]

\subfloat[Peres gate with $T$-depth 3]{
\mbox{
\Qcircuit @C=0.1em @R=.5em @!R{
	&\lstick{A}  &\qw  &\gate{T}  	&\ctrl{1}  	&\targ 	 &\qw  		&\qw 			&\targ  		&\gate{T^\dag} 	&\targ  		&\qw  		&\qw &\qw &\rstick{P=A} \\
	&\lstick{B}   &\qw   &\gate{T} 	&\targ   	&\qw 	&\ctrl{1}    &\gate{T^\dag}	&\ctrl{-1} 	 &\gate{T^\dag} 	&\qw   		&\ctrl{1}  	&\qw &\qw &\rstick{Q=A\oplus B}\\
	&\lstick{C}   &\gate{H} &\gate{T} &\qw 	&\ctrl{-2} 	&\targ		&\qw   			&\qw 	  		&\gate{T} 		&\ctrl{-2}  &\targ  		&\gate{H} &\qw  &\rstick{R=AB\oplus C}\\
	}
}

}

\subfloat[Toffoli gate with $T$-depth 3]{
\mbox{
\Qcircuit @C=0.1em @R=.5em @!R{
	&\ctrl{2} & \qw &  & &\lstick{A}  &\qw  &\gate{T}  &\targ   	&\qw 	&\ctrl{2}   &\qw 		&\ctrl{1} &\gate{T^\dag} &\qw  &\ctrl{2}  &\targ &\qw &\rstick{P=A} \\
	&\ctrl{1} &\qw   		& \push{\rule{.3em}{0em}=\rule{.8em}{0em}}& &\lstick{B}   &\qw   &\gate{T} &\ctrl{-1} 	&\targ 	&\qw   &\gate{T^\dag}	&\targ 	 &\gate{T^\dag} &\targ   &\qw  &\ctrl{-1} &\qw &\rstick{Q=B}\\
	&\targ &\qw   		& & &\lstick{C}   	&\gate{H} &\gate{T} &\qw 	&\ctrl{-1} &\targ	&\qw   		&\qw 	  &\gate{T} 	&\ctrl{-1}  &\targ  &\gate{H} &\qw  &\rstick{R=AB\oplus C}\\
	}
}

}

\subfloat[TR gate with $T$-depth 4]{
\mbox{
\Qcircuit @C=0.1em @R=.5em @!R{
	&\ctrl{1} & \qw    & & &\lstick{A}  &\qw    &\qw   	&\gate{T^\dag}   		&\ctrl{1} 	&\targ 	&\gate{T} &\qw   		&\targ  &\qw &\qw  &\rstick{P=A} \\
	&\ctrlo{1} & \qw &\push{\rule{.0em}{0em}=\rule{.8em}{0em}}  & &\lstick{B}   &\gate{T} &\targ 	&\gate{T^\dag} 		&\targ   	&\qw 	 &\gate{T}   &\targ    &\qw  &\gate{T^\dag}  &\qw &\rstick{Q=A\oplus B}\\
	&\targ &\qw   		& & &\lstick{C} 	&\gate{H} 	&\ctrl{-1} &\gate{T^\dag}	&\qw   		&\ctrl{-2}  &\qw  &\ctrl{-1}    &\ctrl{-2} &\gate{H} &\qw  &\rstick{R=A\overline{B}\oplus C}\\
	}
}
}
\caption{Fault-tolerant architecture for Toffoli gate, Peres gate and TR gate}
\label{fig:fault-tolerant Toffoli, Peres, TR}
\end{figure}
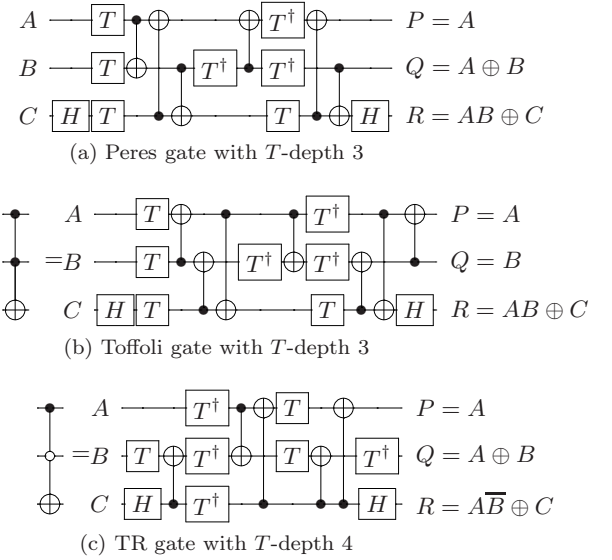 

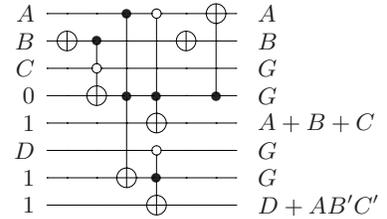
\begin{figure}[H]
\centering
\mbox{
\Qcircuit @C=0.4em @R=0.3em @!R{
	&\lstick{A}     	&\qw   	&\qw 		&\ctrl{3}	&\ctrlo{3} 	&\qw		&\targ	&\qw		&\rstick{A} \\
	&\lstick{B}  	&\targ 	&\ctrl{1} 	&\qw 		&\qw			&\targ 	&\qw		&\qw		&\rstick{B}\\
	&\lstick{C}  	&\qw  	&\ctrlo{1}	&\qw 		&\qw 		&\qw		&\qw		&\qw		&\rstick{G}\\
	&\lstick{0}     	&\qw 	&\targ		&\ctrl{3} 	&\ctrl{1}	&\qw		&\ctrl{-3}	&\qw	&\rstick{G}\\
	&\lstick{1}    	&\qw  	&\qw 		&\qw 		&\targ		&\qw		&\qw 	&\qw		&\rstick{A+B+C}\\
	&\lstick{D}     	&\qw  	&\qw 		&\qw 		&\ctrlo{1} 	&\qw		&\qw		&\qw		&\rstick{G}\\
	&\lstick{1} 		&\qw 	&\qw 		&\targ		&\ctrl{1} 	&\qw		&\qw		&\qw		&\rstick{G}\\
	&\lstick{1}    	&\qw 	&\qw  		&\qw 		&\targ 		&\qw 	&\qw		&\qw	&\rstick{D+AB'C'}\\ 
	} 
}
\caption{Fault-Tolerant RLZC circuit}
\label{fig:ft-RLZC}
\end{figure}

The RLZC is mainly constructed from 3 TR gates and one Toffoli gate. We also propose a fault-tolerant design for RLZC gate by using the fault-tolerant designs of Toffoli gate and TR gate in Fig.~\ref{fig:fault-tolerant Toffoli, Peres, TR}. The design is shown in Fig.\ref{fig:ft-RLZC} and this gate has $T$-depth of 11.

\subsection{Controlled-$V$ and Controlled-$V^\dag$ Gates}

\begin{figure}[h]

\centering
\subfloat[Fault-Tolerant Controlled-$V$ gate]{
\mbox{
\Qcircuit @C=1em @R=.7em @!R{
	& \ctrl{1} & \qw & \push{\rule{.3em}{0em}=\rule{.3em}{0em}} &   &\gate{T} 	&\targ		&\gate{T\dag}	&\targ 		&\qw 		&\qw    \\
	&\gate{V} &\qw   		& &  &\gate{H} 	&\ctrl{-1} 	&\gate{T}  		&\ctrl{-1} 	&\gate{H}	&\qw \\
	}
	}

}

\subfloat[Fault-Tolerant Controlled-$V^\dag$ gate]{
\mbox{
\Qcircuit @C=1em @R=.7em @!R{
	& \ctrl{1} & \qw & \push{\rule{.3em}{0em}=\rule{.3em}{0em}} &   &\qw 		&\targ		&\gate{T}		&\targ 		&\gate{T\dag} 	&\qw  \\
	&\gate{V^\dag} &\qw   		& & &\gate{H} 	&\ctrl{-1} 	&\gate{T\dag}  	&\ctrl{-1} 	&\gate{H}		&\qw \\
	}
	}

}

\caption{Fault-Tolerant architecture for Controlled-$V$ and Controlled-$V^\dag$ gate}
\label{fig:fault-tolerant controlled-V}
\end{figure}
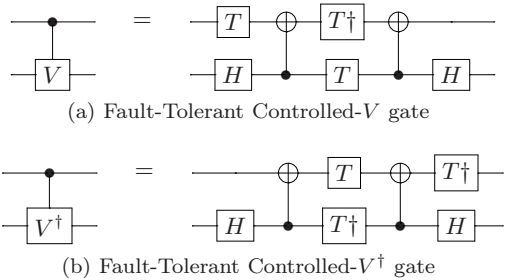

To build the fault-tolerant designs for reversible half subtractor and reversible full subtractor described in previous sections, we incorporate fault-tolerant designs of the Controlled-$V$~\cite{fast-synthesis} and Controlled-$V^\dag$ gates. The decomposition of the Controlled-$V^\dag$ gate may be constructed from Controlled-$V$ by putting adjoint operators of Controlled-$V$ in reverse order. Fig.~\ref{fig:fault-tolerant controlled-V} shows the decomposition of these designs.

\subsection{Reversible Full Adder}

For our 28-bit adder, we use 27 RFA gates (Reversible Full Adder) and a one RHA gate, which is simply implemented by a single Peres gate. Fig.~\ref{fig:fault-tolerant RFA} shows the decomposition of RFA gate described in~\cite{fast-synthesis}. This RFA gate has $T$-depth of 2.

\begin{figure*}[ht]
\centering
\mbox{
\Qcircuit @C=0.4em @R=.5em @!R{
	&\lstick{A}  &\qw   		&\qw  		&\gate{T}		&\ctrl{1}	&\targ 		&\qw			&\ctrl{1} &\gate{T^\dag}	&\ctrl{1}	&\qw			&\targ	&\qw		&\qw	 	&\rstick{P=A} \\
	&\lstick{B}  &\qw   		&\qw  		&\gate{T}		&\targ 		&\qw 		&\ctrl{1}	&\targ	&\gate{T^\dag}	&\targ		&\qw			&\qw		&\qw		&\qw		&\rstick{Q=A\oplus B} \\
	&\lstick{C}  &\qw 		&\ctrl{1}  	&\gate{T}		&\ctrl{1}	&\qw 		&\targ		&\ctrl{1}	&\gate{T^\dag}	&\ctrl{1}	&\qw			&\qw		&\qw		&\qw		&\rstick{R=A\oplus B \oplus C}\\
	&\lstick{D}  &\gate{H} 	&\targ 		&\gate{T^\dag}	&\targ  		&\ctrl{-3}  &\qw 		&\targ	&\gate{T}		&\targ	 	&\gate{P}	&\ctrl{-3}	&\gate{H} &\qw &\rstick{S=(A\oplus B)\oplus C \oplus AB}\\
	}
}
\caption{Circuit implementing a reversible 1-bit Reversible Full Adder.}
\label{fig:fault-tolerant RFA}
\end{figure*}

\subsection{Reversible Half Subtractor and Reversible Full Subtractor}

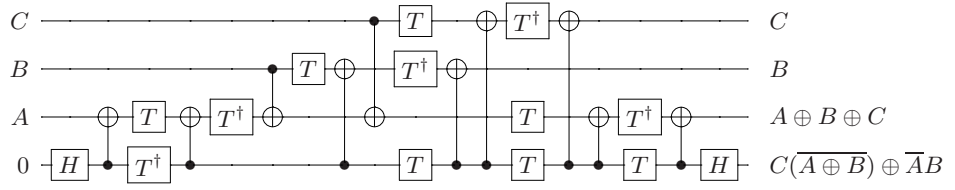
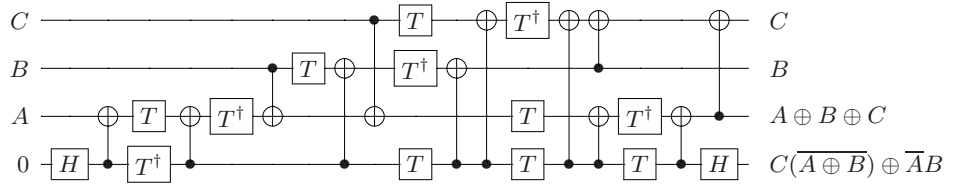
\begin{figure*}[ht]

\subfloat[RFS1 gate corresponds to Fig.~\ref{fig:proposed RFS1} with $T$-depth 6]{
\mbox{
\Qcircuit @C=0.4em @R=.5em@!R{
	&\lstick{C} 	 &\qw 		&\qw 	&\qw				&\qw		&\qw			&\qw	 	  &\qw		&\qw 	&\ctrl{2} &\gate{T} 			&\qw 	&\targ 		&\gate{T^\dag} 	&\targ &\qw &\qw	&\qw &\qw &\qw	&\rstick{C} \\
	&\lstick{B} 	 &\qw 		&\qw 	&\qw				&\qw		&\qw			&\ctrl{1} &\gate{T}	&\targ 		&\qw 	&\gate{T^\dag} 	&\targ 		&\qw 		&\qw 		  &\qw 		&\qw &\qw &\qw &\qw &\qw &\rstick{B}\\
	&\lstick{A}  &\qw 		&\targ 	&\gate{T}		&\targ	&\gate{T^\dag} &\targ &\qw	 	&\qw 		&\targ 	&\qw 			&\qw 		&\qw 		&\gate{T} 		&\qw 	&\targ &\gate{T^\dag}  &\targ &\qw &\qw &\rstick{A\oplus B\oplus C}\\
	&\lstick{0}  &\gate{H} &\ctrl{-1} &\gate{T^\dag} &\ctrl{-1} &\qw		&\qw	 	&\qw	  		&\ctrl{-2} 	&\qw		&\gate{T}		&\ctrl{-2} 	&\ctrl{-3} 	&\gate{T} 		&\ctrl{-3} &\ctrl{-1} &\gate{T} &\ctrl{-1} &\gate{H} &\qw &\rstick{C(\overline{A\oplus B})\oplus \overline{A}B} 
	} 
}
}

\centering

\subfloat[RFS2 gate corresponds to Fig.~\ref{fig:proposed RFS2} with $T$-depth 6]{
\mbox{
\Qcircuit @C=0.4em @R=.5em @!R{
	&\lstick{C} 	 &\qw 		&\qw 	&\qw				&\qw		&\qw			&\qw	 	  &\qw		&\qw 	&\ctrl{2} &\gate{T} 			&\qw 	&\targ 		&\gate{T^\dag} 	&\targ &\targ &\qw	&\qw &\targ &\qw	&\rstick{C} \\
	&\lstick{B} 	 &\qw 		&\qw 	&\qw				&\qw		&\qw			&\ctrl{1} &\gate{T}	&\targ 		&\qw 	&\gate{T^\dag} 	&\targ 		&\qw 		&\qw 		  &\qw 		&\ctrl{-1} &\qw &\qw &\qw &\qw &\rstick{B}\\
	&\lstick{A}  &\qw 		&\targ 	&\gate{T}		&\targ	&\gate{T^\dag} &\targ &\qw	 	&\qw 		&\targ 	&\qw 			&\qw 		&\qw 		&\gate{T} 		&\qw 	&\targ &\gate{T^\dag}  &\targ &\ctrl{-2} &\qw &\rstick{A\oplus B\oplus C}\\
	&\lstick{0}  &\gate{H} &\ctrl{-1} &\gate{T^\dag} &\ctrl{-1} &\qw		&\qw	 	&\qw	  		&\ctrl{-2} 	&\qw		&\gate{T}		&\ctrl{-2} 	&\ctrl{-3} 	&\gate{T} 		&\ctrl{-3} &\ctrl{-1} &\gate{T} &\ctrl{-1} &\gate{H} &\qw &\rstick{C(\overline{A\oplus B})\oplus \overline{A}B} 
	} 
}

}

\caption{Our proposed designs for Fault-Tolerant Reversible Full Subtractor.}
\label{fig:tolerant-full-subtractor}
\end{figure*}

We use the fault-tolerant design of Controlled-$V$ and Controlled-$V^\dag$ gates to incorporate RFS and RHS gates. After eliminating the gates which cancel each other when they are applied to the same qubit, we have the fault-tolerant designs of RFS and RHS gates which are shown in Fig.~\ref{fig:tolerant-half-subtractor} and Fig.~\ref{fig:tolerant-full-subtractor}.

\begin{figure}[h]

\centering

\subfloat[RHS1 gate corresponds to Fig.~\ref{fig:proposed RHS1} with $T$-depth 4]{
\mbox{
\Qcircuit @C=0.25em @R=.5em @!R{
	&\lstick{A}  &\qw    &\qw   	&\gate{T^\dag}   		&\ctrl{1} 	&\targ 	&\gate{T} &\qw   		&\targ  &\qw &\ctrl{1} &\qw &\rstick{P=A} \\
	&\lstick{B}   &\gate{T} &\targ 	&\gate{T^\dag} 		&\targ   	&\qw 	 &\gate{T}   &\targ    &\qw  &\gate{T^\dag} &\targ &\qw &\rstick{Q=B}\\
	&\lstick{0} 	&\gate{H} 	&\ctrl{-1} &\gate{T^\dag}	&\qw   		&\ctrl{-2}  &\qw  &\ctrl{-1}    &\ctrl{-2} &\qw &\gate{H} &\qw  &\rstick{R=A\overline{B}}\\
	}
}
}

\subfloat[RHS2 gate corresponds to Fig.~\ref{fig:proposed RHS2} with $T$-depth 4]{
\mbox{
\Qcircuit @C=0.25em @R=.5em @!R{
	&\lstick{0}  &\qw    &\qw   	 &\qw    &\qw   &\qw    &\qw    &\qw    &\qw   &\qw  &\targ &\qw &\qw &\rstick{R=A\oplus B} \\
	&\lstick{A}  &\qw    &\qw   	&\gate{T^\dag}   		&\ctrl{1} 	&\targ 	&\gate{T} &\qw   		&\targ  &\qw &\qw  &\ctrl{1} &\qw &\rstick{P=A} \\
	&\lstick{B}   &\gate{T} &\targ 	&\gate{T^\dag} 		&\targ   	&\qw 	 &\gate{T}   &\targ    &\qw  &\gate{T^\dag} &\ctrl{-2}  &\targ &\qw &\rstick{Q=B}\\
	&\lstick{0} 	&\gate{H} 	&\ctrl{-1} &\gate{T^\dag}	&\qw   		&\ctrl{-2}  &\qw  &\ctrl{-1}    &\ctrl{-2} &\qw &\gate{H} &\qw &\qw &\rstick{S=A\overline{B}}\\
	}
}
}

\caption{Our proposed designs for Fault-Tolerant Reversible Half Subtractor.}
\label{fig:tolerant-half-subtractor}
\end{figure}

\subsection{Metrics for Fault-Tolerant Quantum Circuit}

In the previous section we evaluated our proposed design in term of quantum cost, garbage output and constant cost in order to directly compare it with prior work. However, in quantum computing we often use KQ ~\cite{fault-tolerant} as the cost metric, which helps to calculate the demands on quantum error correction. KQ is calculated by multiplying the number of qubits used and the circuit depth, or number of time steps. Here we use the fault-tolerant design, thus we use $T$-depth as the circuit's depth. 

\par Table~\ref{tab:T-depth} shows the $T$-depth of each stage in the reversible floating point adder. Therefore, we can evaluation of our proposed design in term of KQ. Note that this total depth is calculated after making some parts run in parallel.

\begin{table}
\centering
\begin{tabular}{|c|c|}
\hline
\textbf{Stage} & \textbf{$T$-depth} \\
\hline
 Swap & 174  \\ \hline
 Alignment & 194  \\ \hline
 Addition &  57 \\ \hline
 Conversion & 212  \\ \hline
 Normalization & 244  \\ \hline
 Rounding & 0 \\ \hline 
 Total &  881 \\ \hline 
\end{tabular}
\\
\caption{$T$-depth of each step in the whole architecture.}
\label{tab:T-depth}
\end{table} 

The total KQ for the whole architecture is 723,301. This compares to a KQ for a 32-bit CDKM ripple-carry adder~\cite{cuccaro04:new-quant-ripple} of 12,474. A floating-point addition is thus nearly sixty times as expensive as fixed-point.
    
\section{Discussion and Conclusion}

With improvements in the two most hardware-intensive parts, this proposed design has reduced the Quantum Cost by 68\%, the Garbage Output by 72\% and the Constant Input by 71.5\%. We also give a fault-tolerant version of the whole architecture.

\par At this stage of the execution, the system state corresponds to $\langle A',B',f(A,B),G \rangle $ where Table~\ref{tab:total_compare} has included $A'$, $B'$ and $G$ under ``garbage output''. To complete the reversibility of the circuit, we must bring in an additional 32-bit register, execute transverse CNOTs from the output value, then run our complete circuit in reverse to clean up all of the garbage as shown in Eq. (1). Thus, the complete circuit uses 821 qubits: 64 variable input qubits and 757 input ancillae. On output, as noted, ancillae are returned to their pristine state, but 32 have been drafted into permanent use. We conclude that floating point addition is not a ``green'' operation, unsustainable with repeated use. 

\par In future work, we plan to investigate restricting the ranges of input values to determine if the reversibility can be improved.
\subsubsection*{Acknowledgments.}

This work was supported by the Japan Society for the Promotion of Science (JSPS) through its ``Funding Program for World-Leading Innovative R\&D on Science and Technology (FIRST Program)''.

\bibliography{fpa}

\end{document}